\documentclass[english]{article}
\usepackage[T1]{fontenc}
\usepackage[latin9]{inputenc}
\usepackage{geometry}
\usepackage{float}
\usepackage{amsmath}
\usepackage{graphicx}
\usepackage{setspace}

\makeatletter

\providecommand{\tabularnewline}{\\}
\usepackage{amsfonts}
\usepackage{amssymb}
\usepackage{cmll}
\DeclareFontFamily{U}{msb}{}
\DeclareFontShape{U}{msb}{m}{n}{ <5> <6> <7> <8> <9> gen * msbm
        <10> <10.95> <12> <14.4> <17.28> <20.74> <24.88> msbm10}{}
\DeclareSymbolFont{AMSb}{U}{msb}{m}{n}
\DeclareMathSymbol{\realset}{\mathalpha}{AMSb}{"52}

\usepackage{babel}
\makeatother

\begin{document}
\noindent \begin{center}
\textbf{\Large Embedding protein $3\mathbf{D}$-structures in a cubic
lattice.}\\
\textbf{\Large{} I. The basic algorithms.}
\par\end{center}{\Large \par}

\vspace{1cm}

\begin{center}
{\Large Jacques Gabarro-Arpa}
\par\end{center}{\Large \par}

\begin{singlespace}
\noindent \begin{center}
\vspace{1cm}
LBPA CNRS, Ecole Normale Sup\'erieure de Cachan
\par\end{center}

\noindent \begin{center}
61, Avenue du Pr\'esident Wilson - 94235 Cachan cedex - France
\par\end{center}

\noindent \begin{center}
e-mail: jga@gtran.org
\par\end{center}
\end{singlespace}

\vspace{2cm}

\noindent \textbf{\Large Abstract}{\Large \par}

\medskip{}

Realistic $3D$-conformations of protein structures can be embedded
in a cubic lattice using exclusively integer numbers, additions, subtractions
and boolean operations.\vspace{1.5cm}

\noindent \textbf{\Large 1. Introduction}{\Large \par}

\medskip{}

In previous papers $\left[1-5\right]$ we have built a series of mathematical
tools for studying the multidimensional molecular conformational space
of biological macromolecules, with the aim of understanding the dynamical
states of proteins by building a complete energy surface $\left[6,7\right]$.

An $N$-atom molecule has a $(N-1)^{3}$-dimensional conformational
space (\textit{CS}), the sheer complexity of this huge structure can
be reduced to tractable dimensions by partitioning it with central
hyperplanes%
\footnote{That pass through the origin.%
} into a finite set of cells, this amounts to discarding all knowledge
about molecular conformations other than the cells that contain them.

In our approach $\left[1\right]$, a set $\mathcal{H}$ of $N_{\mathcal{H}}=N\times(N-1)/2$
hyperplanes generates a partition in \textit{CS} of $N!^{3}$ cells,
on the other hand hyperplanes are oriented structures dividing\textit{
}the space into a $+$ and a $-$ half-spaces, thus points within
a cell are characterized by a binary sequence of length $N_{\mathcal{H}}$
enumerating the orientations with respect the hyperplane set. This
binary sequence is all the information that remains from the molecular
conformations.

Our choice of hyperplanes $\{H_{ij}\in\mathcal{H}:c_{i}-c_{j}=0,\quad0\leq i<j\leq N-1,\quad c\in\{x,y,z\}\}$%
\footnote{A convention used here is that $c$ represents any of the cartesian
coordinates $x$, $y$, $z$.%
} $\left[1\right]$, is such that the $+$/$-$ hemispaces are the
points with $c_{i}>c_{j}$ and $c_{i}<c_{j}$ respectively. This induces
an order relation in the $x$, $y$ and $z$ coordinates of points
in a cell \medskip{}

$c_{\alpha_{0}}<c_{\alpha_{1}}<c_{\alpha_{2}}<...<c_{\alpha_{N-2}}<c_{\alpha_{N-1}}$\hspace*{82.1mm}$(1)$\medskip{}

\noindent where $\{\alpha_{0},\alpha_{1},\alpha_{2},...,\alpha_{N-2},\alpha_{N-1}\}$,
a permutation of the sequence $\{0,1,2,...,N-2,N-1\}$, is the \textbf{dominance
partition sequence} \textit{(DPS)}$\left[1\right]$\textit{.} 

The compactedness and hierarchical structure of the codes generated
by partition sequences made possible the construction of a graph whose
nodes are the cells in \textit{CS} that are visited by the thermalized
molecule with edges towards adjacent cells, this was the subject developped
in previous works $\left[2-5\right]$. 

However interesting this result may be, it is of no practical use
unless on top of it there is a method for calculating the energy of
molecular conformations in a cell. With the mesoscopic force field
approximations currently used in molecular simulations $\left[8,9\right]$,
where atoms are represented as point-like structures, the only input
to the Hamiltonian energy function are the interatomic distances calculated
from $3D$ molecular conformations. In this framework the purpose
of this work is twofold: 

\begin{enumerate}
\item given a partition sequence, we want to calculate a fair sample of
compatible $3D$ molecular conformations, 
\item we want to encode the set of sampled conformations with a combinatorial
structure so they can be more easily manipulated.
\end{enumerate}
\hspace*{4.6mm}In the following sections are described the algorithms
for doing this: 

\begin{itemize}
\item In section 2 we build a complete set of lattice covalent bond segments,
which are the basic building blocks: the whole molecular structure
is built upon them.
\item The \textit{DPS}s can be seen as the lattice projections of a molecular
structure where all intervals in each dimension are reduced to one
lattice spacing (Fig. 5 of $\left[1\right]$), these have to be increased
locally to obtain a realistic structure. In section 3 we build the
partially ordered set of lattice intervals between bonded atoms, a
structure needed for calculating the maximum and minimum expansion
values of each interval, this gives a set of linear inequalities described
in section 4.
\item In section 5 it is shown how an inter-dependent system of inequalities
can be made independent.
\item In section 6 the form and structure of the system of linear inequalities
is discussed in detail.
\end{itemize}
\begin{figure}[H]
\hspace*{6mm}\includegraphics[angle=270,scale=0.7]{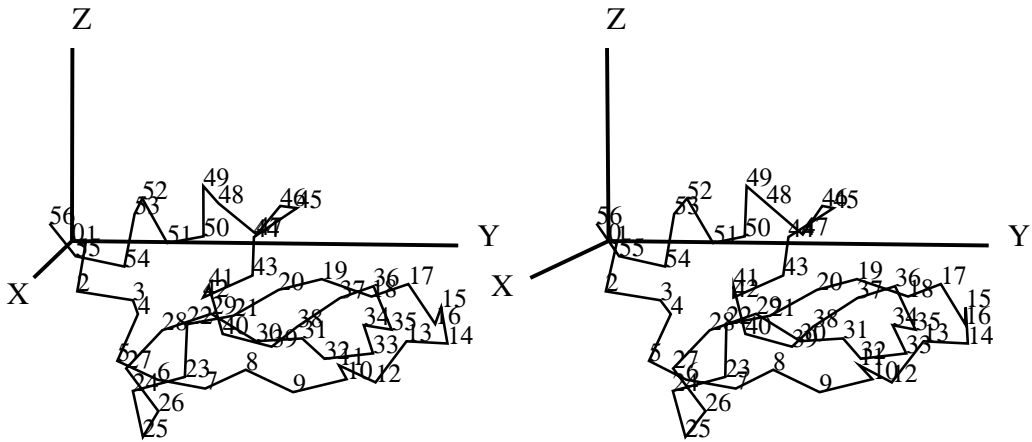}

\caption{Stereoview of a \textit{pancreatic trypsin inhibitor protein} (\textit{PTI})
$C_{\alpha}$-backbone molecular conformation (Table I), corresponding
to the dominance partition sequences in Fig. 2.}

\end{figure}

To illustrate the algorithmic methods that are the subject of the
present work, we have chosen as an example (Fig. 1 and Table I) the
$C_{\alpha}$-backbone of the pancreatic trypsin inhibitor protein
$\left[10\right]$, because it is a small protein molecule and the
mathematical structures it generates are of moderate size, yet it
has the complexity that can be found in longer molecules. Also the
side chains have been put aside for the same reason: they would have
made the contents of Figs. 2 and 3 almost unreadable.

\noindent \begin{flushleft}
\begin{figure}[H]
\hspace*{25mm}\includegraphics[scale=0.6]{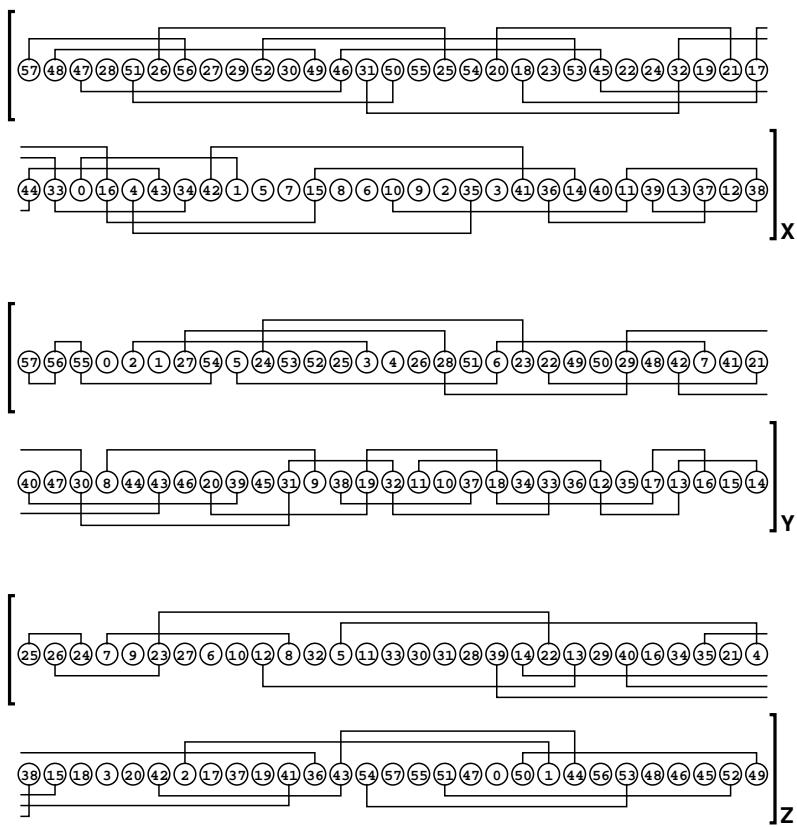}

\caption{Dominance partition sequence of the \textit{PTI} $C_{\alpha}$-backbone
for the molecular conformation from Fig. 1. Showing the maximal intervals
for each coordinate. }

\end{figure}

\par\end{flushleft}

\noindent \begin{flushleft}
\textbf{\Large 2. The expanded lattice covalent bond segments set\medskip{}
}
\par\end{flushleft}{\Large \par}

The numbers in $x$, $y$ and $z$ dominance partition sequences can
be regarded as the evenly spaced projections of $N$ points in a $3D$
cubic lattice, it is a particular form of embedding where the separation
between consecutive projections of atoms in $x$, $y$ and $z$ has
been shrinked to \underbar{one} lattice spacing. The aim of the present
work is to expand this embedding so to obtain realistic molecular
structures.

To do this we must restrict the most basic element of molecular structures:
the covalent bond, to a finite set of coordinate values, such that
with a suitable unit of length can be transformed to give integer
values exclusively. These restricted bonds can still be useful for
describing real molecular conformations if the minimum magnitude of
vector differences is small enough. This can be done, for the example
developped here (\textit{PTI} $C_{\alpha}$-backbone), using empirical
data sampled from molecular dynamics simulations $\left[11\right]$,
it requires the following steps

\begin{enumerate}
\item First we determine the dimensions of the lattice by taking as reference
the mean bond length and its range of variation for bonded $C_{\alpha}$
pairs, in our case this gives: $3.58$\AA< $3.86$\AA< $4.13$\AA.
We set arbitrarily the bond mean length to $20$ lattice units, which
gives a lattice spacing of $0.19$\AA. Thus, any segment between
two lattice points with a length range between $3.58\times20/3.86$
and $4.13\times20/3.86$ is potentially a $C_{\alpha}$-$C_{\alpha}$
bond segment, and the set $B$ of valid \textbf{lattice bond segments},
modulo a lattice translation along the $x$, $y$ and $z$ axes, is
the set of segments starting at the origin and ending in any lattice
point that lies between two spheres of radius $3.58\times20/3.86$
and $4.13\times20/3.86$ respectively. This gives a total of $1883$
primary segments, excluding reflections through the $xy$, $xz$ and
$yz$ planes.
\item Next we determine the range of variation for the bond angles, which
is greater than that for the bond length and varies considerably along
the $C_{\alpha}$chain. For each bond angle $A_{\alpha_{i},\alpha_{i+1},\alpha_{i+2}}$
we determine two integer numbers : the floored minimum $\lfloor min(A_{\alpha_{i},\alpha_{i+1},\alpha_{i+2}})\rfloor$
and the ceiled maximum range $\lceil max(A_{\alpha_{i},\alpha_{i+1},\alpha_{i+2}})\rceil$
respectively. These divide the interval between the absolute minimum
and maximum values $71\textdegree-167\textdegree$ in $64$ subintervals\\
\hspace*{24mm}$71\textdegree$-$74\textdegree$-$75\textdegree$-$76\textdegree$-$77\textdegree$-$78\textdegree$-$79\textdegree$-$80\textdegree$-$81\textdegree$-$82\textdegree$-$87\textdegree$-$89\textdegree$-\\
\hspace*{24mm}$90\textdegree$-$92\textdegree$-$93\textdegree$-$94\textdegree$-$95\textdegree$-$96\textdegree$-$97\textdegree$-$98\textdegree$-$99\textdegree$-$100\textdegree$-$101\textdegree$-\\
\hspace*{24mm}$104\textdegree$-$105\textdegree$-$106\textdegree$-$107\textdegree$-$108\textdegree$-$109\textdegree$-$110\textdegree$-$112\textdegree$-$113\textdegree$-\\
\hspace*{24mm}$114\textdegree$-$115\textdegree$-$116\textdegree$-$117\textdegree$-$118\textdegree$-$119\textdegree$-$120\textdegree$-$121\textdegree$-$124\textdegree$-\\
\hspace*{24mm}$125\textdegree$-$127\textdegree$-$129\textdegree$-$135\textdegree$-$136\textdegree$-$138\textdegree$-$139\textdegree$-$143\textdegree$-$144\textdegree$-\\
\hspace*{24mm}$147\textdegree$-$148\textdegree$-$149\textdegree$-$150\textdegree$-$151\textdegree$-$152\textdegree$-$153\textdegree$-$154\textdegree$-$155\textdegree$-\\
\hspace*{24mm}$156\textdegree$-$157\textdegree$-$159\textdegree$-$162\textdegree$-$163\textdegree$-$167\textdegree$\hspace*{66.3mm}$(2)$
\item The dynamic values of each $A_{\alpha_{i},\alpha_{i+1},\alpha_{i+2}}$
spann a given range of intervals from $(2)$, thus consecutive bonds
$B_{\alpha}$and $B_{\alpha+1}$ can only be assigned discrete bond
segments that form an angle within the specific range.
\end{enumerate}
\hspace*{5.5mm}In building realistic 3D-conformations from the \textit{DPS}s
by embedding these in a bigger lattice, the following problem arises:
the intervals $C_{\alpha_{i}}-C_{\alpha_{i+1}}$between consecutive
$C_{\alpha}$s, for a given coordinate in Fig. 2, must be replaced
by lattice intervals which are generally longer, so the excess lattice
units must be distributed among the intermediate sequence intervals,
such that the resulting lattice segments bonding $C_{\alpha}$s are
from the set of valid lattice bond segments described above.

To solve this problem the following steps are needed

\begin{enumerate}
\item build from the \textit{DPS}s the consecutive $C_{\alpha}$ intervals
poset (Fig. 3), 
\item determine for each consecutive $C_{\alpha}$ interval the maximum
an minimum excess values,
\item make the linear inequalities in $x$, $y$ and $z$ independent of
one another.
\end{enumerate}
\medskip{}

\noindent \textbf{\Large 3. The consecutive $\mathbf{\mathbf{C_{\alpha}}}$
intervals poset\medskip{}
}{\Large \par}

Fig. 2 shows the \textit{DPS}s for the \textit{PTI} $C_{\alpha}$-backbone,
it also shows some of the intervals between consecutive $C_{\alpha}$s
: $\mathcal{I}_{\alpha}^{c}$s %
\footnote{The following naming convention applies to any symbol refering to
a bond interval $C_{\alpha}-C_{\alpha+1}$ : it bears only the smaller
index.%
}, a partial order relation can be defined for them. But first, we
recall some basic definitions : let $\mathcal{I}_{\alpha_{1}}^{c}$
and $\mathcal{\mathcal{I}}_{\alpha_{2}}^{c}$ be two $\mathcal{I}_{\alpha}^{c}$s
spanning the \textit{DPS$_{\text{c}}$} intervals $\left\{ \sigma_{\alpha_{1}}^{c_{left}},\sigma_{\alpha_{1}}^{c_{right}}\right\} $
and $\left\{ \sigma_{\alpha_{2}}^{c_{left}},\sigma_{\alpha_{2}}^{c_{right}}\right\} $

\begin{description}
\item [{Definition\ 1}] $\mathcal{I}_{\alpha_{1}}^{c}$ \textbf{precedes}
$\mathcal{I}_{\alpha_{2}}^{c}$ or $\mathcal{I}_{\alpha_{1}}^{c}$$\prec$
$\mathcal{I}_{\alpha_{2}}^{c}$,\\
 if $\mathcal{I}_{\alpha_{1}}^{c}\subset\mathcal{I}_{\alpha_{2}}^{c}$
or equivalently $\sigma_{\alpha_{1}}^{c_{left}}\geq\sigma_{\alpha_{2}}^{c_{left}}$
and $\sigma_{\alpha_{1}}^{c_{right}}\leq\sigma_{\alpha_{2}}^{c_{right}}$.
\item [{Definition\ 2}] $\mathcal{I}_{\alpha_{2}}^{c}$ \textbf{succeeds}
$\mathcal{I}_{\alpha_{1}}^{c}$or $\mathcal{I}_{\alpha_{2}}^{c}$$\succ$$\mathcal{I}_{\alpha_{1}}^{c}$.
\item [{Definition\ 3}] A\textbf{\emph{ }}\textbf{maximal} interval is
not succeeded by any other interval.
\item [{Definition\ 4}] A \textbf{minimal} interval is not preceeded by
any other interval\emph{.}
\end{description}
Fig. 2 shows the set of maximal intervals for \textit{DPS}$_{\text{\mbox{x}}}$,
\textit{DPS$_{y}$} and \textit{DPS}$_{z}$. 

\begin{description}
\item [{Definition\ 5}] A \textbf{cover} is a set of two intervals $\mathcal{I}_{\alpha_{1}}^{c}$$\prec$
$\mathcal{I}_{\alpha_{2}}^{c}$ with no $\mathcal{I}_{\alpha_{x}}^{c}$such
that $\mathcal{I}_{\alpha_{1}}^{c}$$\prec$$\mathcal{I}_{\alpha_{x}}^{c}$$\prec$$\mathcal{I}_{\alpha_{2}}^{c}$.
\end{description}
\begin{figure}[H]
\hspace*{1mm}\includegraphics[scale=0.75]{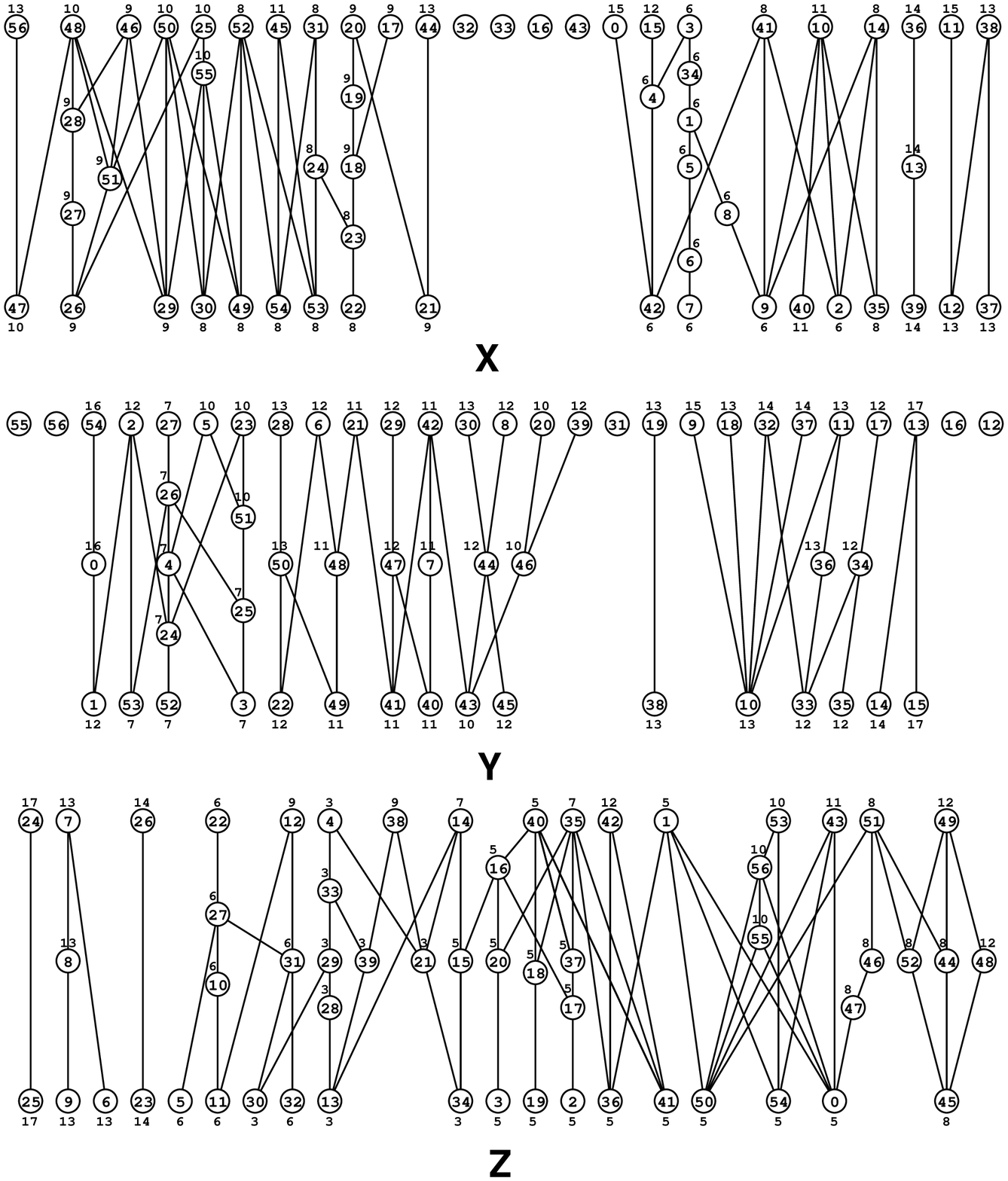}\caption{Consecutive $C_{\alpha}$ intervals cover graph. Minimal/maximal intervals
are at the bottom/top respectively, with succession going from bottom
to top. For each interval }

\end{figure}

Fig. 3 displays a graphical representation of this partially ordered
set (\textbf{poset}), where the nodes are the \textit{$\mathcal{I}_{\alpha}^{c}$}
set and the edge set consists of the pairs satisfying the cover relation.
As we shall see below the poset structure allows to define the set
of linear inequalities for determining the lattice bond segments.
\medskip{}

\noindent \begin{flushleft}
\textbf{\Large 4. Determining the bounds on excess values\medskip{}
}
\par\end{flushleft}{\Large \par}

The excess value of an interval $\mathcal{I}_{\alpha}^{c}$ is the
difference between its length on the \textit{DPS} and on the extended
lattice. In order to expand the \textit{DPS} lattice we must determine
first the bounds of excess values for every $\mathcal{I}_{\alpha}^{c}$.%
\begin{figure}[H]
\hspace*{38mm}\includegraphics[scale=0.7]{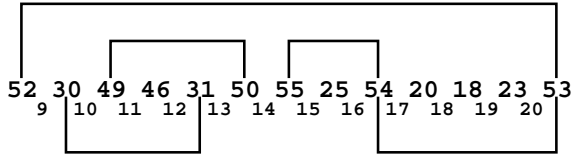}

\caption{Sequence of maximal interval $\mathcal{I}_{52}^{x}$ showing its minimal
preceeding intervals. }

\end{figure}

An example will help to understand, we have in Fig. 4 a set of $5$
connected $\mathcal{I}_{\alpha}^{x}$s: $\mathcal{I}_{52}^{x}$ which
is a maximal interval, and its minimal predecessors $\mathcal{I}_{30}^{x}$,
$\mathcal{I}_{49}^{x}$, $\mathcal{I}_{54}^{x}$ and $\mathcal{I}_{53}^{x}$
(Fig. 3), they fill positions $8$ to $20$ in the $x$-sequence where
the $12$ minimal intervals between $C_{\alpha}$s have local excess
variables $\chi_{9}^{x}$ to $\chi_{20}^{x}$ (Fig. 4), giving the
local expansion value in the extended lattice. The following equations
define the excess values\medskip{}

$X_{52}^{x}=\underset{9\leq\sigma\leq20}{\sum}\chi_{\sigma}^{x}-|\mathcal{I}_{52}^{x}|$
(where the last term is the $c$-sequence interval length) \medskip{}

$X_{30}^{x}=\underset{10\leq\sigma\leq12}{\sum}\chi_{\sigma}^{x}-|\mathcal{I}_{30}^{x}|\qquad\qquad X_{49}^{x}=\underset{11\leq\sigma\leq13}{\sum}\chi_{\sigma}^{x}-|\mathcal{I}_{49}^{x}|$\hspace*{51.8mm}(3)\medskip{}

$X_{54}^{x}=\underset{15\leq\sigma\leq16}{\sum}\chi_{\sigma}^{x}-|\mathcal{I}_{54}^{x}|\qquad\qquad X_{53}^{x}=\underset{17\leq\sigma\leq20}{\sum}\chi_{\sigma}^{x}-|\mathcal{I}_{53}^{x}|$\medskip{}

\noindent also $X_{52}^{x}$ must be greater that the sum of the $X_{\alpha}^{x}$
from preceeding non-overlapping intervals \medskip{}

$X_{52}^{x}\geq X_{49}^{x}+X_{54}^{x}+X_{53}^{x}\qquad\qquad X_{52}^{x}\geq X_{30}^{x}+X_{54}^{x}+X_{53}^{x}$\hspace*{53.2mm}(4)\medskip{}

To build from (4) a complete system of linear inequalities allowing
to calculate the $\chi_{\sigma}^{c}$s for embedding the molecular
system in the extended lattice, first we need to determine the bounds\medskip{}

$Xmin_{\alpha}^{c}\leq X_{\alpha}^{c}\leq Xmax_{\alpha}^{c}$\hspace*{102.9mm}(5)\medskip{}

By construction the maximum lattice bond segment length on any coordinate
is $21$, this gives for the extreme values of excess lattice units
on any interval $\mathcal{I}_{\alpha}^{c}$ the relation \medskip{}

$0\leq|\mathcal{I}_{\alpha}^{c}|+X_{\alpha}^{c}\leq21$ \hspace*{110mm}(6)\medskip{}

\noindent which settles the initial minimum and maximum bond lattice
units for the $c$-coordinate to\medskip{}

$b_{c}^{min}=0\quad$ and $\quad b_{c}^{max}=|\mathcal{I}_{\alpha}^{c}|+21$\hspace*{87mm}(7)\medskip{}

\noindent respectively. Let $B_{c}^{\{b_{c}^{min},b_{c}^{max}\}}$
be the set of all lattice bond segments $b$ such that $b_{c}^{min}\leq b_{c}\leq b_{c}^{max}$
for $c\in\{x,y,z\}$, then the set $B_{\mathcal{I}_{\alpha}}$ of
all the lattice bond segments that are within the bounds (7) is\medskip{}

$B_{\mathcal{I}_{\alpha}}=B_{x}^{\{b_{x}^{min},b_{x}^{max}\}}\cap B_{y}^{\{b_{y}^{min},b_{y}^{max}\}}\cap B_{z}^{\{b_{z}^{min},b_{z}^{max}\}}$
\hspace*{63.7mm}(8)\medskip{}

This operation may change the bounds (7), this is because the $b\in B_{\mathcal{I}_{\alpha}}$
have a common origin but the points at the other extreme form a connected
irregular cluster (see the example in Fig. 5): the bonds excluded
by (8) may be the ones that contain the extremes of other coordinates.
This gives a new set of bonds and the process has to be repeated until
the bounds stabilize.\medskip{}

\hspace*{43.5mm}\includegraphics[scale=0.5]{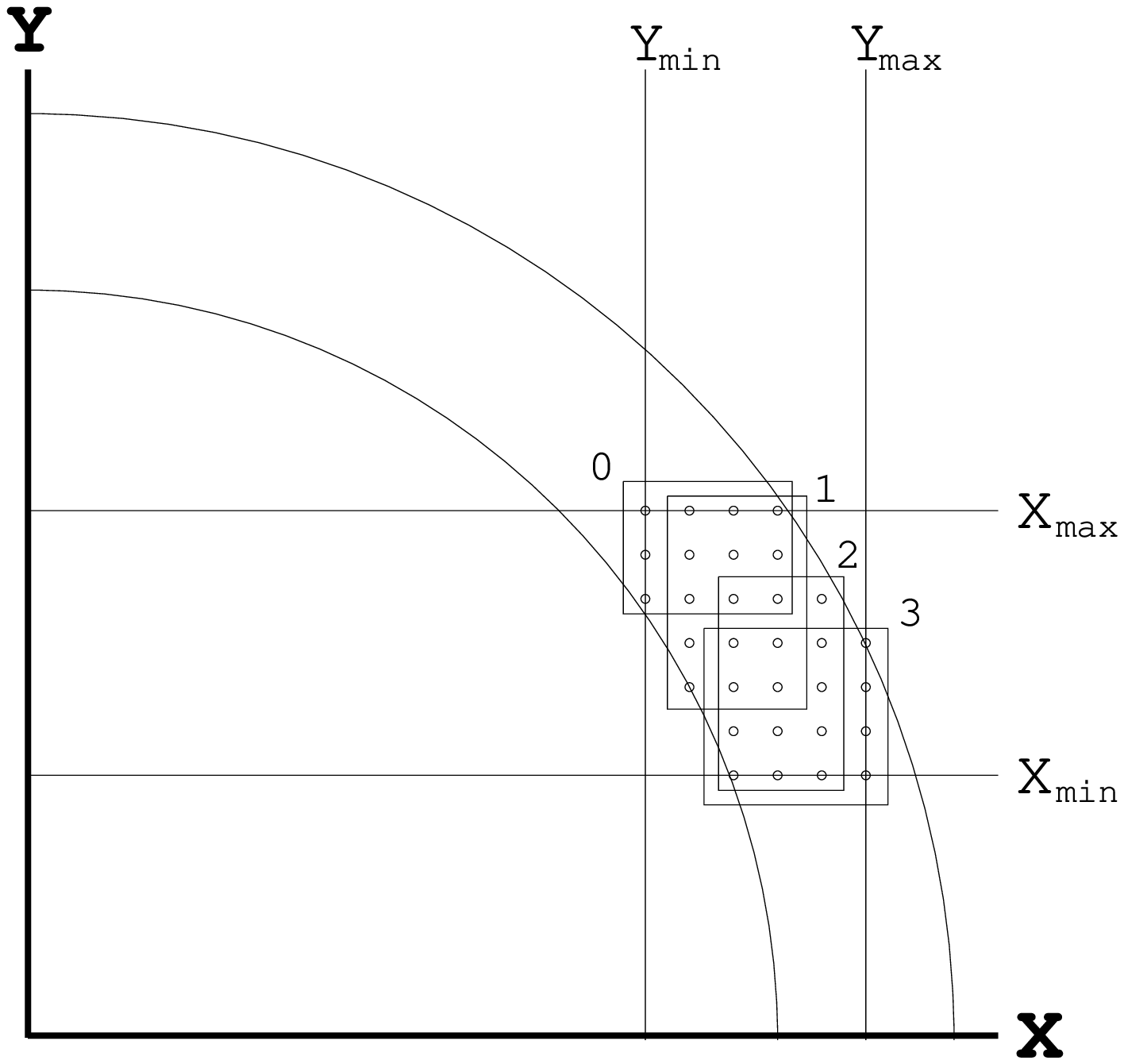}

\begin{figure}[h]
\caption{2D example of a $B_{\mathcal{I}_{\alpha}}$ set. The lattice bond
segments ($b$) start at the origin, and end on any lattice point
in the region bounded by the two spheres described in section $2.1$.\protect \\
The only $b$ end points shown are those lying within the $x$ and
$y$ bounds, or equivalently $b\in B_{\mathcal{I}_{\alpha}}$.\protect \\
As shown in the picture the set $B_{\mathcal{I}_{\alpha}}$ can be
decomposed into the minimal covering set of $n_{p}=4$ rectangular
subsets $\mathcal{P}_{\mathcal{I}_{\alpha}}^{0}$, $\mathcal{P}_{\mathcal{I}_{\alpha}}^{1}$,
$\mathcal{P}_{\mathcal{I}_{\alpha}}^{2}$ and $\mathcal{P}_{\mathcal{I}_{\alpha}}^{3}$.}

\end{figure}
\medskip{}

\noindent \begin{flushleft}
\textbf{\Large 5. Making the inequalities independent \medskip{}
}
\par\end{flushleft}{\Large \par}

From the set of bounds (7) we can build the set of linear inequalities
(using again the example from the previous section)\medskip{}

$Xmax_{52}^{x}\geq\underset{9\leq\sigma\leq20}{\sum}\chi_{\sigma}^{x}\geq Xmin_{52}^{x}$\medskip{}

$Xmax_{30}^{x}\geq\underset{10\leq\sigma\leq12}{\sum}\chi_{\sigma}^{x}\geq Xmin_{30}^{x}\qquad\qquad Xmax_{49}^{x}\geq\underset{11\leq\sigma\leq13}{\sum}\chi_{\sigma}^{x}\geq Xmin_{49}^{x}$\hspace*{25.2mm}(9)\medskip{}

$Xmax_{54}^{x}\geq\underset{15\leq\sigma\leq16}{\sum}\chi_{\sigma}^{x}\geq Xmin_{54}^{x}\qquad\qquad Xmax_{53}^{x}\geq\underset{17\leq\sigma\leq20}{\sum}\chi_{\sigma}^{x}\geq Xmin_{53}^{x}$\medskip{}

There is a further problem to be taken into consideration: $Xmin_{\alpha}^{c}$
and $Xmax_{\alpha}^{c}$ are the $c$-coordinate bounds of the set
$B_{\mathcal{I}_{\alpha}}$ but, due to the non-uniform shape of $B_{\mathcal{I}_{\alpha}}$,
selecting one or more $c$-values in this interval while discarding
the rest may change completely the bounds in the other coordinates.
This the induces an interdependence between inequalities (9) in $x$,
$y$ and $z$, in which case solving the system becomes much more
complex.

This problem can be avoided if the end points of bonds in $B_{\mathcal{I}_{\alpha}}$
fill completely a lattice rectangular parallelopiped, in this case
the choice of bounds in one coordinate leaves the others unchanged.
Thus $B_{\mathcal{I}_{\alpha}}$ has to be decomposed into a set of
rectangular parallelopipeds $\mathbf{P}_{\mathcal{I}_{\alpha}}$ \medskip{}

$B_{\mathcal{I}_{\alpha}}=$ $\underset{0 < p\leq n_p}{\bigcup}$$\mathcal{P}_{\mathcal{I}_{\alpha}}^{p},\quad\mathcal{P}_{\mathcal{I}_{\alpha}}^{p}\in\mathbf{P}_{\mathcal{I}_{\alpha}}$\hspace*{90mm}(10)\medskip{}

\noindent subject to the following conditions

\begin{enumerate}
\item \noindent there are no $\mathcal{P}_{\mathcal{I}_{\alpha}}^{p_{1}}\in\mathbf{P}_{\mathcal{I}_{\alpha}}$
and $\mathcal{P}_{\mathcal{I}_{\alpha}}^{p_{2}}\in\mathbf{P}_{\mathcal{I}_{\alpha}}$
such that $\mathcal{P}_{\mathcal{I}_{\alpha}}^{p_{1}}\subset\mathcal{P}_{\mathcal{I}_{\alpha}}^{p_{2}}$,
\item $n_{p}$ is minimal,
\item for $\mathbf{P}_{\mathcal{I}_{\alpha}}$obeying conditions 1 and 2
and $\mathcal{P}_{\mathcal{I}_{\alpha}}^{p_{1}}\in\mathbf{P}_{\mathcal{I}_{\alpha}}$
there is no $\mathcal{P}_{\mathcal{I}_{\alpha}}^{p_{2}}$ such that
$|\mathcal{P}_{\mathcal{I}_{\alpha}}^{p_{1}}|<|\mathcal{P}_{\mathcal{I}_{\alpha}}^{p_{2}}|$.
\end{enumerate}
\noindent \hspace*{5.5mm}In that case shriking the bounds of a set
$\mathcal{P}_{\mathcal{I}_{\alpha}}^{p}$ for any coordinate does
not alter the bounds in the other dimensions and thus solutions to
the inequalities can be found independently for each coordinate.

\medskip{}

\noindent \begin{flushleft}
\textbf{\Large 6. The structure of the solutions\medskip{}
}
\par\end{flushleft}{\Large \par}

The inequalities (9), for instance, can be rewritten as\medskip{}

$Xmax_{52}\geq U_{9,20}.\chi^{x}\geq Xmin_{52}\quad...$\hspace*{85.2mm}(11)\medskip{}

\noindent where the $U_{\mathcal{I}_{\alpha}^{c}}$s are $(N-1)$-dimensional
vectors of the form\medskip{}

$U_{\mathcal{I}_{\alpha}^{c}}=(0,...,0,1,...,1,0,...,0)$ \hspace*{93mm}(12)\medskip{}

\noindent with ones in the contiguous positions from $\sigma_{\alpha}^{c_{left}}$
to $\sigma_{\alpha}^{c_{right}}$ and zeros everywhere else, and $\chi^{x}$
is the vector\medskip{}

$\chi^{x}=(\chi_{0}^{x},...,\chi_{9}^{x},...,\chi_{20}^{x},...,\chi_{N-1}^{x})$\hspace*{88.2mm}(13)\medskip{}

Extending this notation to the whole set of inequalities for $0\leq\alpha\leq N-1$
and $x$, $y$ and $z$, we have

\medskip{}

$Xmax_{\mathcal{I}_{\alpha}}^{x}\geq U_{\mathcal{I}_{\alpha}^{x}}.\chi^{x}\geq Xmin_{\mathcal{I}_{\alpha}}^{x}$

\medskip{}

$Xmax_{\mathcal{I}_{\alpha}}^{y}\geq U_{\mathcal{I}_{\alpha}^{y}}.\chi^{y}\geq Xmin_{\mathcal{I}_{\alpha}}^{y}$\hspace*{92.3mm}(14)

\medskip{}

$Xmax_{\mathcal{I}_{\alpha}}^{z}\geq U_{\mathcal{I}_{\alpha}^{z}}.\chi^{z}\geq Xmin_{\mathcal{I}_{\alpha}}^{z}$\medskip{}

Taking the vectors $U_{\mathcal{I}_{\alpha}^{c}}$ as the rows of
a $(N-1)\times(N-1)$ matrix $U^{c}$, and $Xmax_{\mathcal{I}_{\alpha}}^{c}/Xmin_{\mathcal{I}_{\alpha}}^{c}$
as the components of vectors $Xmax^{c}/Xmin^{c}$ (14) can be rewritten
as\medskip{}

$Xmax^{x}\geq U^{x}.\chi^{x}\geq Xmin^{x}\,\,\,\,\,\, Xmax^{y}\geq U^{y}.\chi^{y}\geq Xmin^{y}\,\,\,\,\,\, Xmax^{z}\geq U^{z}.\chi^{z}\geq Xmin^{z}$\hspace*{6.8mm}(15)\medskip{}

\noindent The above set of inequalities define $2\times(N-1)$ affine
half-spaces $\mathrm{Hmin_{\mathcal{I_{\alpha}}}^{\mathnormal{c}}}$
and $\mathrm{Hmax_{\mathcal{I_{\alpha}}}^{\mathnormal{c}}}$ whose
intersection determines an $\mathrm{H}$-polytope in \textit{$CS^{c}$}
\textit{$\left[12,13\right]$.} Hence, the vertices of this polytope
are among the unique solutions of the $3\times2^{N-1}$ systems of
equations

\noindent \medskip{}

$U^{x}.\chi^{x}=Xlim^{x}\qquad U^{y}.\chi^{y}=Xlim^{y}\qquad U^{z}.\chi^{z}=Xlim^{z}\qquad0\leq\alpha\leq N-1$\hspace*{22.6mm}(16)\medskip{}

\noindent where $Xlim^{c}$ can be either $Xmax^{c}$ or $Xmin^{c}$
and the $\geq$ relation in (15) has been restricted to $=$. Moreover,
the matrices $U^{c}$ with rows like (12) are called \textbf{interval
matrices}, they belong to a very important class of matrices called:
\textbf{totally unimodular matrices} $\left[12\right]$. These have
the particularity that the determinant of any minor is either $-1$,
$0$ or 1. This ensures that the vertices of the polytope are integer
vectors (or lattice points), since solving (16) by applying the Cramer's
rule the denominator is always $-1$ or $1$. Thus, the solutions
of (16) can be written

\medskip{}

$\chi^{c}=\overline{U}^{c}.Xlim^{c}$\hspace*{116mm}(17)\medskip{}

\noindent where $\overline{U}^{c}$ is the inverse of $U^{c}$.

The $\mathrm{V}$-polytope is the representation of the polytope by
its set of vertices, these can be obtained from (17) by determining
the combinations in $Xlim^{c}$ compatible with (15). The solutions
of the system of linear inequalities (15) can be generated from this
set through convex combinations, as the three sets of inequalities
are independent the general solution will be the product of the $x$,
$y$ and $z$ polytopes.

The total unimodularity of matrix $U^{c}$ also ensures that most
combinatorial algorithms can be run in polynomial time.

\noindent \medskip{}

\begin{flushleft}
\textbf{\Large \newpage{}7. Conclusion\medskip{}
}
\par\end{flushleft}{\Large \par}

The purpose of the line of work being developped here, is to show
that molecular structures can be built and analysed with a fraction
of the information (in our case less than $1/5$) that can be found
in a typical PBD file.

This might seem a significant but modest quantitative difference,
but qualitatively is more than that: discarding information results
in the emergence of mathematical structures that were buried in the
complexity of the data, which in turn can be encoded efficiently by
them. Using combinatorics a great number of molecular conformations
can be dealt simultaneously, thus overcoming the barrier that computations
have to be performed on the basis of one conformation at a time.

The algorithmic method developped before $\left[1-5\right]$ serves
two purposes 

\begin{enumerate}
\item As an \textbf{amplifier} : by codifying data sampled in computer simulations
into discrete gemetrical structures, these can be combined to generate
an estimate of the volume occupied by a molecule in its conformational
space.
\item As a molecular 3D-structure \textbf{compressor} : it is possible to
translate basic features of molecular 3D-structures into a binary
code, which in turn can be very efficiently amalgamated into ternary
sequences that encode great numbers of cells from \textit{CS}. The
information on the whole \textit{CS} volume can be cast into a file
compatible with desktop memory size.
\end{enumerate}
\hspace*{5.5mm}The present work is the first one of a third and last
step: the development of combinatorial methods for calculating the
energy of structures from cells in \textit{CS}. 

Here we have developped the basic algorithms for this : realistic
discrete protein conformations can be built and embedded in a cubic
lattice, using a table of discrete bond segments and, more important,
these conformations can be encoded into combinatorial structures.

However many issues still remain unexplored:

\begin{itemize}
\item \begin{flushleft}
The possible combinations of $\mathcal{P}_{\mathcal{I}_{\alpha}}^{p}$s
from (10) is a huge set, efficient sampling methods should be developped.
\par\end{flushleft}
\item The $\mathrm{V}$-polytope should be better characterized.
\item The present formalism should be extended to take into account sets
of adjacent cells.
\item Last of all inter-atomic distances should also be encoded into combinatorial
structures.
\end{itemize}
\begin{flushleft}
These will be dealt in forthcoming works.\textbf{\Large \newpage{}8.
Appendix\medskip{}
}
\par\end{flushleft}{\Large \par}

\begin{description}
\item [{Table}] 1. Lattice coordinates of the \textit{PTI} $C_{\alpha}$-backbone
from Fig. 1.
\item [{Column}] \ \ \ \ \ \ $\alpha$ \ \ \ \ \ \ : $C_{\alpha}$
number.
\item [{Columns}] $x_{\alpha}$\ $y_{\alpha}$\ $z_{\alpha}$\ \ :
$C_{\alpha}$ coordinates.
\item [{Columns}] $b_{x}$\ \ $b_{y}$\ \ $b_{z}$ \ : bond vector
between $C_{\alpha-1}$ and $C_{\alpha}$.
\end{description}
\noindent \begin{flushleft}
\begin{tabular}{|c|ccccc|ccccc||c|ccccc|ccccc|}
\hline 
$\alpha$ &  & $x_{\alpha}$ & $y_{\alpha}$ & $z_{\alpha}$ &  &  & $b_{x}$ & $b_{y}$ & $b_{z}$ &  & $\alpha$ &  & $x_{\alpha}$ & $y_{\alpha}$ & $z_{\alpha}$ &  &  & $b_{x}$ & $b_{y}$ & $b_{z}$ & \tabularnewline
\hline
\hline 
0 &  & 0 & 0 & 0 &  &  &  &  &  &  & 29 &  & -34 & 49 & -30 &  &  & 6 & 19 & 7 & \tabularnewline
\hline 
1 &  & 19 & 7 & 1 &  &  & 19 & 7 & 1 &  & 30 &  & -33 & 66 & -40 &  &  & 1 & 17 & -10 & \tabularnewline
\hline 
2 &  & 30 & 5 & -16 &  &  & 11 & -2 & -17 &  & 31 &  & -25 & 84 & -38 &  &  & 8 & 18 & 2 & \tabularnewline
\hline 
3 &  & 31 & 26 & -19 &  &  & 1 & 21 & -3 &  & 32 &  & -9 & 94 & -44 &  &  & 16 & 10 & -6 & \tabularnewline
\hline 
4 &  & 12 & 26 & -26 &  &  & -19 & 0 & -7 &  & 33 &  & -1 & 113 & -41 &  &  & 8 & 19 & 3 & \tabularnewline
\hline 
5 &  & 19 & 19 & -43 &  &  & 7 & -7 & -17 &  & 34 &  & 14 & 111 & -29 &  &  & 15 & -2 & 12 & \tabularnewline
\hline 
6 &  & 28 & 35 & -49 &  &  & 9 & 16 & -6 &  & 35 &  & 30 & 123 & -29 &  &  & 16 & 12 & 0 & \tabularnewline
\hline 
7 &  & 19 & 52 & -53 &  &  & -9 & 17 & -4 &  & 36 &  & 38 & 117 & -12 &  &  & 8 & -6 & 17 & \tabularnewline
\hline 
8 &  & 27 & 68 & -45 &  &  & 8 & 16 & 8 &  & 37 &  & 55 & 106 & -15 &  &  & 17 & -11 & -3 & \tabularnewline
\hline 
9 &  & 29 & 86 & -53 &  &  & 2 & 18 & -8 &  & 38 &  & 64 & 91 & -25 &  &  & 9 & -15 & -10 & \tabularnewline
\hline 
10 &  & 28 & 106 & -48 &  &  & -1 & 20 & 5 &  & 39 &  & 50 & 80 & -34 &  &  & -14 & -11 & -9 & \tabularnewline
\hline 
11 &  & 47 & 105 & -41 &  &  & 19 & -1 & 7 &  & 40 &  & 44 & 61 & -30 &  &  & -6 & -19 & 4 & \tabularnewline
\hline 
12 &  & 60 & 120 & -46 &  &  & 13 & 15 & -5 &  & 41 &  & 36 & 55 & -13 &  &  & -8 & -6 & 17 & \tabularnewline
\hline 
13 &  & 54 & 131 & -31 &  &  & -6 & 11 & 15 &  & 42 &  & 17 & 51 & -19 &  &  & -19 & -4 & -6 & \tabularnewline
\hline 
14 &  & 40 & 145 & -33 &  &  & -14 & 14 & -2 &  & 43 &  & 12 & 69 & -11 &  &  & -5 & 18 & 8 & \tabularnewline
\hline 
15 &  & 24 & 141 & -21 &  &  & -16 & -4 & 12 &  & 44 &  & -3 & 68 & 2 &  &  & -15 & -1 & 13 & \tabularnewline
\hline 
16 &  & 6 & 137 & -29 &  &  & -18 & -4 & -8 &  & 45 &  & -14 & 83 & 12 &  &  & -11 & 15 & 10 & \tabularnewline
\hline 
17 &  & -3 & 126 & -15 &  &  & -9 & -11 & 14 &  & 46 &  & -32 & 75 & 11 &  &  & -18 & -8 & -1 & \tabularnewline
\hline 
18 &  & -15 & 111 & -21 &  &  & -12 & -15 & -6 &  & 47 &  & -44 & 65 & -1 &  &  & -12 & -10 & -12 & \tabularnewline
\hline 
19 &  & -8 & 93 & -14 &  &  & 7 & -18 & 7 &  & 48 &  & -50 & 50 & 10 &  &  & -6 & -15 & 11 & \tabularnewline
\hline 
20 &  & -16 & 76 & -19 &  &  & -8 & -17 & -5 &  & 49 &  & -33 & 46 & 18 &  &  & 17 & -4 & 8 & \tabularnewline
\hline 
21 &  & -7 & 60 & -28 &  &  & 9 & -16 & -9 &  & 50 &  & -24 & 47 & 0 &  &  & 9 & 1 & -18 & \tabularnewline
\hline 
22 &  & -13 & 42 & -32 &  &  & -6 & -18 & -4 &  & 51 &  & -38 & 32 & -4 &  &  & -14 & -15 & -4 & \tabularnewline
\hline 
23 &  & -15 & 41 & -52 &  &  & -2 & -1 & -20 &  & 52 &  & -34 & 23 & 13 &  &  & 4 & -9 & 17 & \tabularnewline
\hline 
24 &  & -10 & 22 & -57 &  &  & 5 & -19 & -5 &  & 53 &  & -15 & 22 & 9 &  &  & 19 & -1 & -4 & \tabularnewline
\hline 
25 &  & -18 & 25 & -75 &  &  & -8 & 3 & -18 &  & 54 &  & -17 & 18 & -11 &  &  & -2 & -4 & -20 & \tabularnewline
\hline 
26 &  & -36 & 29 & -67 &  &  & -18 & 4 & 8 &  & 55 &  & -24 & -1 & -8 &  &  & -7 & -19 & 3 & \tabularnewline
\hline 
27 &  & -35 & 17 & -51 &  &  & 1 & -12 & 16 &  & 56 &  & -36 & -12 & 3 &  &  & -12 & -11 & 11 & \tabularnewline
\hline 
28 &  & -40 & 30 & -37 &  &  & -5 & 13 & 14 &  & 57 &  & -52 & -15 & -9 &  &  & -16 & -3 & -12 & \tabularnewline
\hline
\end{tabular}\\

\par\end{flushleft}

\noindent \begin{flushleft}
\textbf{\Large \newpage{}References\medskip{}
}
\par\end{flushleft}{\Large \par}

\begin{description}
\item [{$\left[1\right]$}] \textbf{A central partition of molecular conformational
space. I. Basic structures}.\\
J. Gabarro-Arpa,\textit{ Comp. Biol. \& Chem.}  \textbf{27}, 153-159
(2003).
\item [{$\left[2\right]$}] \textbf{A central partition of molecular conformational
space. II. Embedding 3D structures}.\\
J. Gabarro-Arpa, \textit{Proceedings of the 26th Annual International
Conference of the IEEE EMBS, San Francisco}, 3007-3010 (2004).
\item [{$\left[3\right]$}] \textbf{A central partition of molecular conformational
space. III. Combinatorial determination of the volume spanned by a
molecular system in conformational space}.\\
J. Gabarro-Arpa, \textit{J. Math. Chem.} \textbf{42}, 691-706 (2006).
\item [{$\left[4\right]$}] \textbf{A central partition of molecular conformational
space. IV. Extracting information from the graph of cells}.\\
J. Gabarro-Arpa, \textit{J. Math. Chem.} \textbf{44}, 872-883 (2006).
\item [{$\left[5\right]$}] \textbf{A central partition of molecular conformational
space. V. The Hypergraph of 3D Partition Sequences}.\\
 J. Gabarro-Arpa \texttt{}~\\
\texttt{arXiv:0812.2844} (2008).
\item [{$\left[6\right]$}] \textbf{Energy landscapes}.\\
D.J. Wales, Cambridge University Press, ISBN 0-521-81415-4, (2003).
\item [{$\left[7\right]$}] \textbf{Dynamic personalities of proteins}.\\
K. Henzler-Wildman, D. Kern, \textit{Nature} \textbf{450}, 964-972
(2007).
\item [{$\left[8\right]$}] \textbf{All-atom empirical potential for molecular
modeling and dynamics studies of proteins}.\\
A.D. MacKerell Jr., et al., \textit{J. Phys. Chem. B}\textbf{ 102},
3586\textendash{}3616 (1998).
\item [{$\left[9\right]$}] \textbf{Biomolecular simulations: recent developments
in force fields, simulations of enzyme catalysis, protein-ligand,
protein-protein, and protein-nucleic acid noncovalent interactions}.\\
W. Wang, O. Donini, C.M. Reyes, P.A. Kollman\\
\textit{Annu. Rev. Biophys. Biomol. Struct.} \textbf{30}, 211\textendash{}243
(2001).
\item [{$\left[10\right]$}] \textbf{The geometry of the reactive site
and of the peptide groups in trypsin, trypsinogen and its complexes
with inhibitors}.\\
M. Marquart, J. Walter, J. Deisenhofer, W. Bode, R. Huber\\
\textit{Acta Crystallogr. Sect. B} \textbf{39}, 480\textendash{}490
(1983).
\item [{$\left[11\right]$}] \textbf{Clustering of a molecular dynamics
trajectory with a Hamming distance}.\\
J. Gabarro-Arpa, R. Revilla,\textit{ Comput. Chem.} \textbf{24}, 693-698
(2000).
\item [{$\left[12\right]$}] \textbf{Theory of Linear and Integer Programming}.\\
 A. Schrijver, John Wiley \& sons, ISBN 0-471-98232-6, pp. 155-156
(1998).
\item [{$\left[13\right]$}] \textbf{Lattice points and lattice polytopes}.\\
 A. Barvinok, \textit{Hanbook of Discrete and Computational Geometry},
CRC Press, ISBN 0-8493-8524-5, pp. 133-152 (1997).
\end{description}

\end{document}